\documentclass[11pt,letterpaper]{article}
\usepackage[margin=1in]{geometry}
\usepackage[T1]{fontenc}
\usepackage{lmodern}
\usepackage{amsmath,amssymb,amsthm}
\usepackage{booktabs,array}
\usepackage{graphicx}
\usepackage{tikz}
\usetikzlibrary{arrows.meta,positioning}
\usepackage[font=small,labelfont=bf]{caption}
\usepackage{microtype}
\usepackage[colorlinks=true,linkcolor=blue,citecolor=blue,urlcolor=blue]{hyperref}
\hypersetup{pdftitle={Shared Phase Arithmetic for Parallel Quantum Rotations},pdfauthor={Anbang Wu; Liqiang Lu; Pu Pang; Jianwei Yin; Jingwen Leng; Minyi Guo},pdfsubject={Parallel phase kickback and logical gate complexity}}
\newcommand{\ket}[1]{\lvert #1\rangle}
\newcommand{\bra}[1]{\langle #1\rvert}

\newcommand{\wt}{\operatorname{wt}}
\newcommand{\supp}{\operatorname{supp}}
\newcommand{\PG}{\ket{G_b}}
\newcommand{\Add}{\operatorname{ADD}_b}
\newcommand{\eps}{\varepsilon}
\newtheorem{theorem}{Theorem}
\newtheorem{proposition}{Proposition}

\title{Shared Phase Arithmetic for Parallel Quantum Rotations}
\author{%
\begin{tabular}{ccc}
Anbang Wu$^{1}$ & Liqiang Lu$^{2}$ & Pu Pang$^{1}$\\[0.2em]
\small\href{mailto:anbang@cs.sjtu.edu.cn}{\texttt{anbang@cs.sjtu.edu.cn}} &
\small\href{mailto:liqianglu@zju.edu.cn}{\texttt{liqianglu@zju.edu.cn}} &
\small\href{mailto:pangpu@cs.sjtu.edu.cn}{\texttt{pangpu@cs.sjtu.edu.cn}}\\[0.7em]
Jianwei Yin$^{2}$ & Jingwen Leng$^{1}$ & Minyi Guo$^{1}$\\[0.2em]
\small\href{mailto:zjuyjw@cs.zju.edu.cn}{\texttt{zjuyjw@cs.zju.edu.cn}} &
\small\href{mailto:leng-jw@cs.sjtu.edu.cn}{\texttt{leng-jw@cs.sjtu.edu.cn}} &
\small\href{mailto:guo-my@cs.sjtu.edu.cn}{\texttt{guo-my@cs.sjtu.edu.cn}}\\[0.5em]
\multicolumn{3}{c}{\small $^{1}$Shanghai Jiao Tong University}\\
\multicolumn{3}{c}{\small $^{2}$Zhejiang University}
\end{tabular}}
\date{}

\begin{document}
\maketitle
\vspace{-1.6em}
\begin{abstract}
A layer of diagonal quantum gates can be specified by an integer-valued function on computational basis states. This description suggests implementing the layer by evaluating that function reversibly and translating its value into a phase using a shared quantum register. We develop this viewpoint for parallel phase kickback (PPK): the weighted sum of several rotation parameters is computed coherently, added once to a phase-gradient state, and then uncomputed. The construction separates the cost of representing a phase function from the cost of applying it. We prove correctness, give explicit error bounds, and analyze logical gate counts for general weights, pairs of rotations, and weights with disjoint binary support. With Clifford-only encoding, a batch uses at most $4(b-1)$ T gates at $b$-bit phase resolution, excluding initialization. The first phase reference is prepared with independent rotations synthesized by Gridsynth; additional phase-gradient states then follow from the disjoint-support construction with $O(b)$ T gates each. The resulting bounds identify when shared phase arithmetic reduces amortized rotation cost and when encoding overhead limits that advantage.
\end{abstract}

\section{Introduction}\label{sec:intro}

Diagonal operations admit two different descriptions: as a sequence of elementary rotations, or as a function assigning a phase to each computational basis state. The first description leads naturally to gate-by-gate approximation. The second exposes arithmetic structure that can be shared across a collection of gates. For example, a product of $M$ single-qubit phase gates assigns the phase $\sum_j x_j\theta_j$ to a basis string $x$. Although there are $2^M$ possible strings, this phase is specified by only $M$ weights. The relevant algorithmic question is how cheaply that weighted sum can be evaluated and converted into a quantum phase.

In the Clifford+T model, approximating each generic rotation separately gives a familiar reference cost. Ancilla-free synthesis typically uses $3\log_2(1/\eps)$ T gates per rotation, up to lower-order terms~\cite{ross2016}. This estimate concerns individual generic angles and a particular resource model; it is not a lower bound for every angle or for a circuit supplied with auxiliary resource states. A parallel phase-kickback construction instead pays for reversible arithmetic, workspace, and initialization of a phase reference. Its benefit must be assessed by counting these resources together.

Phase kickback provides the conversion from arithmetic to phase. A Fourier eigenstate of modular addition acquires a known phase when shifted, while its quantum state is otherwise unchanged~\cite{kitaev2002,jones2012}. Consequently, a register holding a basis-dependent integer can imprint the corresponding phase with one addition. Reversible evaluation followed by kickback and uncomputation is an established construction. Existing work has used reusable Fourier states for fault-tolerant rotations and for efficient quantum Fourier transforms~\cite{jones2014,nam2020}; low-T adders make this approach especially relevant~\cite{gidney2018}. These results motivate a common description of the phase function, its reversible evaluation, and its subsequent use.

Here we study \emph{parallel phase kickback} (PPK) through precisely this decomposition. Given integer rotation weights $K_j$ at resolution $2^{-b}$, an encoder computes $F(x)=\sum_j x_jK_j\bmod 2^b$ into an auxiliary register. One register addition couples that value to a phase-gradient state. A decoder clears the auxiliary information. The resulting action is the desired product of phase gates. The PPK construction shares the phase-imprinting operation across the target rotations; it does not imply constant circuit depth or free evaluation of $F$.

This formulation yields several concrete resource statements. A general basis-state construction has a worst-case gate cost of $O(Mb2^M)$. Specific features of the $M$ weights can greatly reduce this cost. Two arbitrary weights have a simplified encoding that uses one temporary logical AND. For weights with disjoint binary support, the sum is an XOR and the encoder uses only CNOT gates. In that case, all non-Clifford cost during a batch comes from a single adder: at most $4(b-1)$ T gates with the construction of Ref.~\cite{gidney2018}. A constant amortized T count follows when the number of compatible rotations grows proportionally to the active phase width. It does not follow for unrestricted angle sets.

Initialization is a separate part of this accounting. The first phase-gradient state is prepared by independent $R_z$ rotations synthesized with Gridsynth. Once that reference is available, further phase-gradient states can be prepared using the disjoint-support construction, at a linear T cost in the register width. The seed cost is therefore paid once and can be shared across subsequent preparations and rotations.

The rest of the paper develops the phase conventions and cost model, states the arithmetic problem, and proves the PPK construction together with its complexity bounds. The analysis concerns logical circuits and asymptotic gate growth. A Fourier-transform specialization illustrates the effect of controlled phases and recovers the known favorable scaling of phase-gradient methods; no benchmark-dependent claims are needed.

\section{Phase Arithmetic Preliminaries}\label{sec:prelim}

\subsection{Phase gates and accuracy}
We use
\begin{equation}
 P(\theta)=\operatorname{diag}(1,e^{i\theta})
 =e^{i\theta/2}R_z(\theta).
\end{equation}
Products of $P$ gates and the corresponding products of $R_z$ gates therefore differ only by a global phase. This convention also avoids discarding a relative phase when discussing controlled operations: a controlled-$P(\theta)$ applies $e^{i\theta}$ only to $\ket{11}$, and is not interchangeable with controlled-$R_z(\theta)$ without an additional phase correction.

Let $N=2^b$. For each angle choose the nearest integer representative
\begin{equation}\label{eq:round}
 K_j=\operatorname{round}\!\left(\frac{N\theta_j}{2\pi}\right)\bmod N,
 \qquad \widetilde\theta_j=\frac{2\pi K_j}{N}.
\end{equation}
Let $\Delta_j\in[-\pi,\pi]$ denote the signed difference $\theta_j-\widetilde\theta_j$ modulo $2\pi$. Nearest rounding gives $|\Delta_j|\leq\pi 2^{-b}$. Hence
\begin{equation}\label{eq:error}
 \left\|\bigotimes_{j=0}^{M-1}P(\theta_j)
       -\bigotimes_{j=0}^{M-1}P(\widetilde\theta_j)\right\|
 \leq\sum_j|\Delta_j|
 \leq M\pi 2^{-b}.
\end{equation}
The norm is the operator norm. The first inequality follows either by telescoping the product or by comparing its diagonal entries. For a whole-layer rounding budget $\delta_{\mathrm{ang}}$, it suffices to use
\begin{equation}\label{eq:precision}
 b\geq\left\lceil\log_2\frac{\pi M}{\delta_{\mathrm{ang}}}\right\rceil.
\end{equation}
Exactly representable dyadic angles need no rounding error. Throughout, comparisons at fixed precision specify whether the error applies to one rotation or to an entire computation.

\subsection{A reusable eigenstate of addition}
The $b$-qubit phase-gradient state is
\begin{equation}\label{eq:gradient}
 \PG=\frac{1}{\sqrt{N}}\sum_{y=0}^{N-1}e^{-2\pi i y/N}\ket{y}.
\end{equation}
All integer additions below are modulo $N$. For the shift $S_k\ket{y}=\ket{y+k\bmod N}$, changing the summation variable gives
\begin{equation}\label{eq:eigen}
 S_k\PG=e^{2\pi i k/N}\PG.
\end{equation}
For example, a shift controlled on $q=1$ acts as
\begin{equation}
 (a\ket{0}+c\ket{1})\PG
 \longmapsto a\ket{0}\PG+c\ket{1}S_k\PG
 =\bigl(a\ket{0}+ce^{2\pi i k/N}\ket{1}\bigr)\PG.
\end{equation}
The reference register is unchanged and can be used again. More generally, the register adder
\begin{equation}\label{eq:adder}
 \Add:\ket{s}\ket{y}\longmapsto\ket{s}\ket{y+s\bmod N}
\end{equation}
imprints $e^{2\pi i s/N}$ on every component of a superposition over $s$ when the second input is $\PG$.

\subsection{Logical resource model}
We count T and $T^\dagger$ gates equally, report Clifford gates and workspace separately, and allow computational-basis preparation, measurement, and classical feedforward. The measurement-assisted ripple-carry adder of Ref.~\cite{gidney2018} supplies an implementation of Eq.~\eqref{eq:adder} with
\begin{equation}\label{eq:addcost}
 T_{\mathrm{add}}(b)=4(b-1),\qquad
 C_{\mathrm{add}}(b)=O(b),\qquad D_{\mathrm{add}}(b)=O(b),
\end{equation}
and at most $b-1$ carry qubits in addition to the two operands. These are implementation costs, not lower bounds; fixed bits can simplify an adder. A conventionally controlled ripple-carry adder has $8b+O(1)$ T cost~\cite{gidney2018}. This is a useful reference implementation, rather than a universal lower bound for single-rotation kickback.

The $4$-T temporary logical AND in the same work stores $xy$ in a clean auxiliary qubit. If its logical value is preserved, it can later be erased with an X-basis measurement and a measurement-conditioned Clifford correction. This is the source of the explicit constants used below. An arbitrary nonlinear encoder does not automatically have a zero-T decoder.

\section{The Phase-Accumulation Problem}\label{sec:problem}

Consider data $\ket{\psi}=\sum_x\alpha_x\ket{x}$ on $M$ qubits. The target operation is
\begin{equation}\label{eq:target}
 U_{\boldsymbol\theta}\ket{x}
 =\exp\!\left(i\sum_{j=0}^{M-1}x_j\theta_j\right)\ket{x}.
\end{equation}
At finite precision, define its integer phase function
\begin{equation}\label{eq:function}
 F_{\boldsymbol K}(x)=\left(\sum_{j=0}^{M-1}x_jK_j\right)\bmod 2^b.
\end{equation}
The computational task is to implement $e^{2\pi i F_{\boldsymbol K}(x)/2^b}$ while returning the workspace to a reusable state.

\subsection{Coherent evaluation rather than classical loading}
A suitable encoder satisfies
\begin{equation}\label{eq:encoder}
 E_F:\ket{x}\ket{0}_S\ket{0}_W
 \longmapsto\ket{x}\ket{F(x)}_S\ket{g(x)}_W,
\end{equation}
where $S$ is a $b$-qubit shift register and $W$ holds any temporary information. The input amplitudes $\alpha_x$ are arbitrary; no uniform superposition over the data is assumed. In particular, preparing a superposition of the individual constants $K_j$ alone would not implement Eq.~\eqref{eq:target}. The shift register must encode their basis-dependent \emph{subset sum}. It has $b$ qubits, independent of the number of possible sums.

The encoder can be realized in multiple ways. The general construction in Section~\ref{sec:theory}.\ref{sec:general} enumerates all $2^M$ input strings and coherently writes the corresponding $b$-bit phase sums. Its worst-case gate cost is $O(Mb2^M)$. When the weights have suitable structure, the same function can be encoded much more cheaply; Section~\ref{sec:theory}.\ref{sec:disjoint} gives a construction using only CNOT gates. The cost therefore depends on both the weights and the chosen realization of $F$.

\subsection{Separating the sources of cost}
Write $T_E$, $T_D$, and $T_{\mathrm{prep}}$ for encoding, decoding, and phase-reference preparation costs. For a single batch, PPK has the achievable cost
\begin{equation}\label{eq:costmaster}
 T_{\mathrm{batch}}=T_E+4(b-1)+T_D.
\end{equation}
For $B$ batches reusing one reference, the total is
\begin{equation}\label{eq:total}
 T_{\mathrm{total}}=T_{\mathrm{prep}}
   +\sum_{r=1}^{B}\bigl[T_{E,r}+4(b_r-1)+T_{D,r}\bigr],
\end{equation}
where $b_r$ may be an active subregister width at most $b$. A batch needs the shift and reference registers, carry workspace, and any encoder workspace. The basic full-width construction uses at most $3b-1+a_E$ auxiliary qubits beyond the data, where $a_E$ is the encoder workspace retained during addition. Parallel batches would need separate resources; reuse here is sequential.

At single-rotation tolerance $\eps$, a gate-by-gate comparison gives approximately $3M\log_2(1/\eps)$ T gates for generic angles~\cite{ross2016}. PPK saves T gates when its full encoding and decoding costs, together with its allocated share of initialization, fit below this reference cost. For an entire layer of error at most $\delta$, both methods instead use a per-angle allowance of order $\delta/M$. The next section makes the resulting trends explicit.

\section{Parallel Phase Kickback (PPK) Theory}\label{sec:theory}

\subsection{Construction and correctness}
\begin{figure}[t]
\centering
\begin{tikzpicture}[x=1cm,y=1cm,>=Latex,font=\small]
 \node[anchor=east] at (0,0) {$Q:\ket{\psi}$};
 \node[anchor=east] at (0,-0.95) {$S:\ket{0}$};
 \node[anchor=east] at (0,-1.9) {$W:\ket{0}$};
 \node[anchor=east] at (0,-2.85) {$G:\PG$};
 \foreach \y in {0,-0.95,-1.9,-2.85} {\draw[->] (0,\y)--(10.6,\y);}
 \draw[fill=blue!7] (0.7,0.35) rectangle (2.7,-2.25);
 \node[align=center] at (1.7,-0.95) {Encode\\$E_F$};
 \draw[fill=blue!7] (7.1,0.35) rectangle (9.1,-2.25);
 \node[align=center] at (8.1,-0.95) {Decode\\$E_F^{-1}$};
 \draw[fill=blue!13] (4.4,-2.53) rectangle (5.9,-3.17);
 \node at (5.15,-2.85) {$+S$};
 \draw (5.15,-0.95)--(5.15,-2.53);
 \fill (5.15,-0.95) circle (2pt);
 \node[fill=white,inner sep=2pt] at (3.45,-0.95) {$\ket{F(x)}$};
 \node[fill=white,inner sep=2pt] at (6.35,-0.95) {$\ket{F(x)}$};
 \node[anchor=west] at (10.65,0) {$U_{\widetilde{\boldsymbol\theta}}\ket{\psi}$};
 \node[anchor=west] at (10.65,-0.95) {$\ket{0}$};
 \node[anchor=west] at (10.65,-1.9) {$\ket{0}$};
 \node[anchor=west] at (10.65,-2.85) {$\PG$};
\end{tikzpicture}
\caption{Parallel phase kickback. Each horizontal wire denotes a register: $Q$ holds the $M$ data qubits; $S$ holds the $b$-bit phase sum; $W$ is optional encoder workspace of $a_E$ qubits; and $G$ holds the $b$-qubit phase-gradient reference $\PG$. The labels $\ket{0}$ denote all-zero registers, not necessarily single qubits. Addition imprints a basis-dependent phase; decoding clears $S$ and $W$. Up to $b-1$ additional adder carry qubits are omitted and are distinct from $W$. A valid measurement-assisted decoder can replace the inverse unitary.}
\label{fig:ppk}
\end{figure}
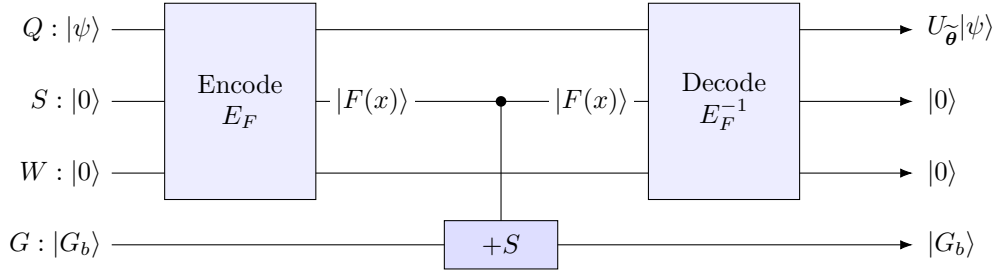

The algorithm has three steps: apply $E_F$, add the shift register to the phase-gradient register, and clear the information computed by $E_F$. Figure~\ref{fig:ppk} shows the dependence of these operations.

During encoding, $S$ coherently stores $F(x)=\sum_jx_jK_j\bmod 2^b$ for each data basis string $x$, while $W$ retains intermediate values $g(x)$, such as a temporary AND. The $+S$ block denotes modular register addition, Eq.~\eqref{eq:adder}, rather than a single-qubit controlled gate. Decoding clears $S$ and $W$; $Q$ retains the desired relative phases and the ideal reference $G$ remains in $\PG$.

\begin{theorem}[Phase-accumulation identity]\label{thm:ppk}
Let $E_F$ satisfy Eq.~\eqref{eq:encoder}, and let $F$ be Eq.~\eqref{eq:function}. On a clean workspace and an exact $\PG$, encoding, addition, and inverse encoding implement $\bigotimes_j P(2\pi K_j/2^b)$. The shift register, workspace, and phase reference return to their input states.
\end{theorem}
\begin{proof}
The three steps act as follows:
\begin{align}
 &\sum_x\alpha_x\ket{x}\ket{0}_S\ket{0}_W\PG\nonumber\\
 &\xrightarrow{E_F}\sum_x\alpha_x\ket{x}\ket{F(x)}_S\ket{g(x)}_W\PG\nonumber\\
 &\xrightarrow{\Add}\sum_x\alpha_xe^{2\pi iF(x)/2^b}
     \ket{x}\ket{F(x)}_S\ket{g(x)}_W\PG\nonumber\\
 &\xrightarrow{E_F^{-1}}\sum_x\alpha_xe^{2\pi iF(x)/2^b}
     \ket{x}\ket{0}_S\ket{0}_W\PG.\label{eq:proof}
\end{align}
Reduction of $F(x)$ modulo $2^b$ changes the exponent by an integer multiple of $2\pi i$. Its phase therefore factors as $\prod_j e^{2\pi i x_jK_j/2^b}$. This is precisely the diagonal action of the stated product of gates. Linearity proves the result also for data entangled with an external system.
\end{proof}

The same argument works for any efficiently computable integer phase function $F$. In particular, replacing $x_j$ by a Boolean predicate $f_j(x)$ implements a product of conditional phases. The predicate evaluation and its erasure must then be included in $T_E$ and $T_D$. An arbitrary diagonal operator specified by $2^M$ unrelated phases may require an exponentially large description and encoder; Theorem~\ref{thm:ppk} does not remove that cost.

\subsection{General weights construction}\label{sec:general}
For arbitrary $M$ weights, a direct construction enumerates the $2^M$ data basis strings. For each string $x$, compute the classical value $F_{\boldsymbol K}(x)$ and apply an $M$-controlled bit flip to each position where that value is one, with the controls selecting $x$. On a zeroed shift register, these operations coherently implement Eq.~\eqref{eq:encoder}. There are at most $b2^M$ such bit flips. Decomposing each into $O(M)$ elementary Clifford+T gates with $O(M)$ reusable clean ancillas gives the worst-case encoding and decoding costs
\[
 T_E+T_D=O(Mb2^M),\qquad C_E+C_D=O(Mb2^M).
\]
The shared phase-imprinting adder contributes only $O(b)$ further gates, so the full construction also has worst-case gate cost $O(Mb2^M)$, excluding reference preparation. Its auxiliary space is $O(M+b)$ when the controlled operations are performed sequentially. These are upper bounds for this explicit construction, not lower bounds on weighted-sum evaluation.

For two rotations, a smaller exact construction is available without a large arithmetic encoder. For data bits $x,y$, define
\begin{equation}
 \Delta=K_0\oplus K_1\oplus\bigl((K_0+K_1)\bmod 2^b\bigr).
\end{equation}
Then the following bit-string identity holds on all four inputs:
\begin{equation}\label{eq:pair}
 F(x,y)=xK_0\oplus yK_1\oplus(xy)\Delta.
\end{equation}
Here $xK$ denotes either the bit string $K$ or zero. CNOTs controlled by $x$ and $y$ load the first two terms. If $\Delta\ne0$, a $4$-T temporary AND computes $t=xy$, and CNOTs from $t$ load the remaining term. Keep $t$ until after addition, undo the loading CNOTs, and erase $t$ by X-basis measurement and an outcome-conditioned CZ on $x,y$. This decoder uses no T gates. Consequently,
\begin{equation}\label{eq:paircost}
 T_{\mathrm{pair}}\leq 4+4(b-1)=4b,
 \qquad T_{\mathrm{pair}}/2\leq 2b.
\end{equation}
The Clifford count is $O(b)$ and the extra retained predicate is one qubit. If $\Delta=0$, the temporary AND is unnecessary. Equation~\eqref{eq:paircost} excludes preparation, as does Eq.~\eqref{eq:costmaster}, and should be compared at the same error allocation as the separate rotations.

More generally, specific features of the $M$ weights can greatly reduce the construction cost. Section~\ref{sec:theory}.\ref{sec:disjoint} develops the case of disjoint binary support, where encoding and decoding require no T gates and the batch cost is only $O(b)$.

\subsection{Disjoint binary support}\label{sec:disjoint}
The strongest simple amortization occurs when no binary position belongs to more than one weight.

\begin{proposition}[Clifford-only encoding]\label{prop:disjoint}
If $\supp(K_j)\cap\supp(K_k)=\varnothing$ for $j\ne k$, then $F_{\boldsymbol K}$ can be encoded and decoded with only CNOT gates. If $h=\sum_j\wt(K_j)$, these two steps together use $2h\leq2b$ CNOTs. A full-width PPK batch then uses at most $4(b-1)$ T gates.
\end{proposition}
\begin{proof}
There are no overlapping one-bits, so no carries occur when any subset of the weights is added. Therefore $\sum_jx_jK_j=\bigoplus_jx_jK_j$. For every one-bit of $K_j$, apply a CNOT from data qubit $j$ to that position of a zeroed shift register. This computes the XOR with $h$ CNOTs. Repeating the CNOTs clears it after addition. The adder supplies the only T gates.
\end{proof}

For nonzero weights this condition implies $M\leq b$. If all weights are divisible by $2^v$, addition never changes the lowest $v$ bits of the phase-gradient register and no carry enters from them. Use the upper $w=b-v$ bits only. By Eq.~\eqref{eq:gradient}, those bits form $\ket{G_w}$; the low bits are an independent factor. Thus the same construction has cost $4(w-1)$, uses at most $2w$ encoding and decoding CNOTs, and requires at most $3w-1$ active auxiliary qubits. A preallocated full reference may still occupy $b$ qubits.

If the binary supports of multiple rotation weights overlap, partition the rotations into $L$ groups so that the weights within each group have pairwise disjoint support. Overlap between different groups is allowed. Apply Proposition~\ref{prop:disjoint} to each group in sequence, reusing the phase-gradient reference and workspace after each group is decoded. Since the phase gates commute, the product of these group operations implements all the desired rotations. If group $\ell$ uses active width $w_\ell\leq b$ as defined above, its T cost is at most $4(w_\ell-1)$, giving
\[
 T_{\mathrm{groups}}\leq4\sum_{\ell=1}^{L}(w_\ell-1)
 \leq4L(b-1),
\]
excluding the one-time reference preparation. Zero-weight rotations can be omitted. Thus overlapping weights can still benefit from disjoint-support encoding, with the total cost determined by the number of groups and their active widths.

For example, $\pi/8$, $\pi/16$, and $\pi/32$ correspond at $b=6$ to weights $4$, $2$, and $1$. Their supports are disjoint, so one $6$-bit addition and six encoding/decoding CNOTs suffice; the adder has $20$ T gates. This is an explicit circuit count, without an empirical claim about a synthesis tool. Special exact gates should always be treated separately in a comparison: $P(\pi/4)$ itself is just T.

For a regular family $\theta_r=2\pi/2^r$, $r=1,\ldots,w$, the weights occupy distinct positions. The construction gives
\begin{equation}\label{eq:fourlimit}
 \frac{T_{\mathrm{batch}}}{M}=\frac{4(w-1)}{w}\longrightarrow 4
 \qquad (M=w\longrightarrow\infty).
\end{equation}
The first few gates in this family have simpler exact implementations and can be removed without changing this limiting trend. Equation~\eqref{eq:fourlimit} concerns increasingly fine, structured angle families with a supplied reference. With $M$ fixed and $w$ growing, the same count grows as $4w/M$ instead of remaining constant.

\subsection{Phase-gradient state preparation and reuse}\label{sec:prep}
The first $b$-qubit phase-gradient state is prepared from $\ket{+}^{\otimes b}$ by independently applying $R_z(-2\pi 2^k/2^b)$ to qubit $k$, for $k=0,\ldots,b-1$. These single-qubit rotations can be synthesized using Gridsynth~\cite{ross2016}; their product gives $\PG$ up to a global phase. Allocating synthesis error $\eta/b$ to each rotation bounds the seed preparation error by $\eta$, with a one-time cost of $O(b\log(b/\eta))$ T gates.

Once this initial state is available, prepare a fresh register in $\ket{+}^{\otimes b}$ and apply PPK using the initial state as the phase reference. The weights $K_k=2^k$ have disjoint binary support, so Proposition~\ref{prop:disjoint} applies directly. A final $X$ gate on each fresh qubit gives the negative-phase convention of Eq.~\eqref{eq:gradient}. Thus each additional phase-gradient state costs at most $4(b-1)$ T gates and $O(b)$ Clifford gates, while preserving the ideal initial reference for further use.

\subsection{Gate-count growth and classical complexity}\label{sec:growth}
Table~\ref{tab:costs} collects the leading counts. Initialization is displayed separately so that the same reference is neither omitted nor charged repeatedly. The direct-synthesis row is for generic angles; exact Clifford or T rotations can be cheaper.

\begin{table}[t]
\centering
\small
\renewcommand{\arraystretch}{1.24}
\begin{tabular}{@{}p{0.31\linewidth}p{0.29\linewidth}p{0.31\linewidth}@{}}
\toprule
Construction & T-count per batch & Conditions and other resources\\
\midrule
Separate synthesis & $\approx 3M\log_2(1/\eps)$ & Generic angles; no shared reference\\
Controlled-adder kickback & $M[8b+O(1)]$ & Reference implementation; $O(b)$ reused workspace\\
PPK, general construction & $O(Mb2^M)$ & Worst-case basis-state encoding; $O(Mb2^M)$ Clifford gates\\
PPK, two weights & $\leq4b$ & One temporary AND; $O(b)$ Clifford gates\\
PPK, disjoint support & $\leq4(w-1)$ & $M\leq w$ for nonzero weights; $O(w)$ Clifford gates and workspace\\
Initial phase reference & $O(b\log(b/\eta))$ & Seed only; independent Gridsynth rotations\\
Additional phase reference & $\leq4(b-1)$ & Reuses the seed; Proposition~\ref{prop:disjoint}\\
\bottomrule
\end{tabular}
\caption{Logical gate growth. Here $\eps$ is a per-rotation synthesis tolerance, $b$ is the reference width, $w\leq b$ is the active width, and $\eta$ is the preparation error. Rotation-implementation costs exclude reference preparation. Measurement and feedforward are allowed.}
\label{tab:costs}
\end{table}

At fixed per-angle tolerance, increasing a compatible batch size reduces the shared adder contribution as $1/M$ until the bit-support constraint is reached. At fixed batch size, tightening the tolerance increases $b$ logarithmically, and the shared contribution grows at the same rate. For a fixed whole-layer budget $\delta$, $b=O(\log(M/\delta))$; the rounding precision and the encoder cost must both change with the problem size. No constant-per-rotation conclusion for general $M$ follows from dividing an isolated adder cost by $M$.

The depth and Clifford resources also matter. The disjoint-support encoder has at most $w$ CNOTs and at worst $O(w)$ depth, followed by an $O(w)$-depth ripple-carry addition and decoding. Thus $O(1)$ amortized T count in Eq.~\eqref{eq:fourlimit} does not mean $O(1)$ circuit depth. The general basis-state construction has worst-case $O(Mb2^M)$ depth and gate count when its controlled operations are executed sequentially. These bounds describe the stated constructions rather than optimal costs for the underlying problem.

The classical construction cost follows the same distinction. Computing the phase sums for all $2^M$ inputs and emitting the general circuit takes $O(Mb2^M)$ elementary construction steps, with output gate count of the same order in the worst case. For the structured case, testing disjoint support and producing its CNOT encoder take only $O(Mb)$ bit operations by scanning the input weights, without enumerating the basis states. This accounting assumes the discretized weights are already supplied; numerical angle evaluation and single-qubit synthesis are separate classical tasks.

\subsection{A Fourier-transform specialization}
The QFT illustrates the gate-count advantage of dyadic rotation rows. We use a representation in which single-qubit phase-rotation rows are separated by Clifford operations, as in phase-gradient QFT constructions~\cite{park2024}. This representation makes the disjoint-support structure explicit. The row cost below refers to these phase-rotation rows after the QFT decomposition.

Consider a row with $m$ rotations of magnitudes $\pi/2^r$, $r=1,\ldots,m$, and let $w=m+1$. Its integer phase function has the form
\begin{equation}\label{eq:qftrow}
 F(x)=\sum_{r=1}^{m}x_r2^{w-r-1}.
\end{equation}
The weights occupy distinct binary positions, so Proposition~\ref{prop:disjoint} gives Clifford-only encoding and decoding. Rows with the opposite phase sign use the inverse phase operation at the same T cost. Thus an achievable row cost is
\begin{equation}\label{eq:qftcost}
 T_{\mathrm{row}}(m)=4m.
\end{equation}
The $4m$ term accounts for the shared phase-imprinting addition; encoding, decoding, and the intervening Clifford operations contribute no T gates.

For an $n$-qubit QFT, the main rows have decreasing lengths, represented in the leading-order count by $m_j=n-j$, $j=1,\ldots,n$. Consequently, their average T cost is
\begin{equation}\label{eq:qftaverage}
 \overline{T}_{\mathrm{row}}
 =\frac{1}{n}\sum_{j=1}^{n}4(n-j)
 =2(n-1)\approx2n.
\end{equation}
Summing over the rows gives
\begin{equation}\label{eq:qfttotal}
 T_{\mathrm{QFT}}
 =4\sum_{j=1}^{n}(n-j)+O(n)+T_{\mathrm{prep}}
 =2n^2+O(n)+T_{\mathrm{prep}}.
\end{equation}
Here the $O(n)$ term accounts for boundary phase rows and end effects of the chosen QFT decomposition. Excluding the separately accounted reference preparation, the leading T-count estimate is therefore $2n^2$. Short rows cost proportionally less than long rows; assigning the maximum row cost to all $n$ rows would overestimate the total.

For an approximate QFT, retain only controlled-phase distances up to $d$. The sum of the omitted phase magnitudes is at most $\pi n2^{-d}$. For $n\geq2$, choosing
\begin{equation}
 d=\min\!\left\{n-1,\left\lceil\log_2\frac{\pi n}{\delta_{\mathrm{cut}}}\right\rceil\right\}
\end{equation}
limits the omission error to $\delta_{\mathrm{cut}}$; no rotation is omitted when $d=n-1$. The resulting phase rows have width $O(d)$ and can share a phase-gradient reference of that width.

The gate-growth bounds are
\begin{align}\label{eq:qftgrowth}
 T_{\mathrm{QFT}}&=O(nd)+O\!\left(d\log\frac{d}{\eta}\right),\\
 C_{\mathrm{QFT}}&=O(nd)+O\!\left(d\log\frac{d}{\eta}\right),\qquad
 A_{\mathrm{QFT}}=O(d),\nonumber
\end{align}
where the additive synthesis term prepares the initial reference and the output-state error is at most $\delta_{\mathrm{cut}}+\eta$. This recovers the established $O(n\log(n/\delta))$ growth of phase-gradient QFT constructions~\cite{nam2020,park2024}, while the untruncated triangular row count gives the $2n^2$ leading term above.

\subsection{Extension to multi-qubit diagonal synthesis}\label{sec:diagonal}
The framework extends directly to an $M$-qubit diagonal operation. After removing an irrelevant global phase, write the target as
\[
 D=\sum_{x\in\{0,1\}^M}e^{i\phi_x}\ket{x}\bra{x},\qquad \phi_{0^M}=0.
\]
Choose $b$-bit shift constants $K_x=\operatorname{round}(2^b\phi_x/2\pi)\bmod 2^b$. The encoder now prepares $\ket{K_x}$ in the shift register for each data basis state $\ket{x}$, replacing the weighted subset sum used for rotations. The same addition and decoding steps then implement the discretized diagonal operation by the argument of Theorem~\ref{thm:ppk}. Only the shift constants loaded during coherent encoding change; the phase-gradient reference and the phase-imprinting adder are unchanged.

For $M=2$, assign the shifts $K_0$ and $K_1$ to $\ket{10}$ and $\ket{01}$, respectively, with zero shift for $\ket{00}$. For two independent rotations, the shift on $\ket{11}$ is constrained to be $(K_0+K_1)\bmod 2^b$. A general two-qubit diagonal operation instead permits an independent shift $K_2$ on $\ket{11}$. Thus the three shifts change from $K_0,K_1,K_0+K_1$ to $K_0,K_1,K_2$, with all values understood modulo $2^b$ and no requirement that $K_2=K_0+K_1$. The encoder in Eq.~\eqref{eq:pair} still applies with $\Delta=K_0\oplus K_1\oplus K_2$, retaining the $4b$ T-count upper bound of Eq.~\eqref{eq:paircost}.

An $M$-qubit diagonal block can contain $2^M-1$ independent relative phases, so implementing the entire block with one phase-imprinting addition may offer greater T-count savings than batching $M$ rotations. Such savings depend on the encoder: without relations among the shift constants, the simplifications available for rotation weights may disappear. The general construction of Section~\ref{sec:theory}.\ref{sec:general} still applies, with worst-case cost $O(Mb2^M)$. Finding inexpensive encoders for arbitrary diagonal blocks is therefore more demanding.

\subsection{Implications and limits}
PPK separates reversible evaluation of a phase function from its conversion to a phase. The separation is useful when the weighted sum has a cheap encoder: pairs admit a constant-size nonlinear component, while disjoint binary supports remove that component entirely. It is less useful when substantial arithmetic is needed to construct every batch. The phase-gradient reference adds a modest, explicitly budgeted startup cost under reuse; it is neither an uncharged resource nor an obstacle that must be repaid for every gate. Together, these facts explain the gate-count trends without assuming universal savings for arbitrary angles or equating low T count with low execution depth.

\begingroup
\small
\bibliographystyle{unsrt}
\bibliography{references}
\endgroup
\end{document}